\documentclass[a4paper,11pt]{article}
\pdfoutput=1
\usepackage{pos}
\usepackage{pgf}
\usepackage{multirow}

\title{Exploring $B$-physics anomalies at colliders}

\author*[a,b]{Jorge Alda}
\author[c]{Jaume Guasch}
\author[a]{Siannah Pe\~naranda}

\affiliation[a]{Departamento de F{\'\i}sica Te{\'o}rica and 
  Centro de Astropart{\'\i}culas y F{\'\i}sica de Altas Energ{\'\i}as (CAPA),\\
  Universidad de Zaragoza,
  Pedro Cerbuna 12,  E-50009 Zaragoza, Spain}

\affiliation[b]{Dipartimento di Fisica e Astronomia ``G. Galilei'', Universit{\`a} degli
Studi di Padova e\\ Istituto Nazionale Fisica Nucleare, Sezione di Padova,
I-35131 Padova, Italy}

        \affiliation[c]{Deptartament de F{\'\i}sica Qu{\`a}ntica i Astrof{\'\i}sica and
  Institut de Ci{\`e}ncies del Cosmos (ICCUB),\\
Universitat de Barcelona, Mart{\'\i} i Franqu{\`e}s 1, E-08028
Barcelona, Catalonia, Spain}

\emailAdd{jalda@unizar.es}
\emailAdd{jaume.guasch@ub.edu, siannah@unizar.es}

\abstract{Several experimental measurements of $B$ meson decays, in tension
  with Standard Model predictions, exhibit large sources of Lepton
  Flavour Universality violation.
  We perform an analysis of the effects of the global fits to the Wilson
  coefficients assuming a model independent effective Hamiltonian
  approach, by including a proposal of different scenarios to include
  the New Physics contributions. Both the current fits at the LHC and
  the ILC projections are considered. We found that for a simultaneous analysis of
predictions for the \RDp and \RKp observables, the scenarios with
three non-universal Wilson coefficients are favoured.}

\FullConference{%
  *** The European Physical Society Conference on High Energy Physics (EPS-HEP2021), ***\\
  *** 26-30 July 2021 ***\\
  *** Online conference, jointly organized by Universität Hamburg and the research center DESY ***
}

\note{This work was partially supported by Spanish Grants
MINECO/FEDER FPA2015-65745-P, PGC2018-095328-B-I00 
(FEDER/Agencia estatal de investigaci{\'o}n) and DGIID-DGA No. 2015-E24/2 (Arag\'on
goverment), Grant No. CB 5/21 (Programa Ibercaja-CAI) and
by MICIN under projects PID2019-105614GB-C22 and 
CEX2019-000918-M of ICCUB (\textit{Unit of Excellence Mar{\'\i}a de Maeztu 2020-2023})
and AGAUR (2017SGR754). J.~A. thanks the warm hospitality of the
Universit\`a degli Studi di Padova and INFN during the completion of this work.}

\newcommand{\RDp}{\ensuremath{ R_{D^{(*)}}} }
\newcommand{\RKp}{\ensuremath{ R_{K^{(*)}}} }

\newcommand{\Clq}{\ensuremath{ C_{\ell q} } }
\newcommand{\Clqt}{\ensuremath{ C_{\ell q(3)} } }
\newcommand{\Clqo}{\ensuremath{ C_{\ell q(1)} } }
\begin{document}
\maketitle
There are several experimental hints of Lepton Flavour Universality Violation (LFUV) in
$b\to c \ell \nu$ and $b\to s \ell^+ \ell^-$ transitions that could be a
sign for physics beyond the Standard
Model (SM). In the first processes,
the $\RDp^\ell$ and $\RDp^\mu$ ratios, defined by
\begin{equation}
\RDp^\ell = \frac{\mathrm{BR}(B \to D^{(*)} \tau \bar{\nu}_\tau ) }{[\mathrm{BR}(B \to D^{(*)} e \bar{\nu}_e) + \mathrm{BR}(B \to D^{(*)} \mu \bar{\nu}_\mu)]/2}\ , \qquad
\RDp^\mu = \frac{\mathrm{BR}(B \to D^{(*)} \tau \bar{\nu}_\tau ) }{ \mathrm{BR}(B \to D^{(*)} \mu \bar{\nu}_\mu)}\ ,
\end{equation}
exhibit a $1.4\,\sigma$ and $2.5\,\sigma$ discrepancy with respect to the
SM predictions, being $3.08\,\sigma$ when combined together. 
The experimental averages for these ratios are~\cite{Amhis:2019ckw}: 
$R_D^\mathrm{ave} = 0.340 \pm 0.027 \pm 0.013, \, R_{D^*}^\mathrm{ave} =
0.295 \pm 0.011 \pm 0.008$. For $b\to s \ell^+ \ell^-$ transitions,
the signs of LFUV are present in the \RKp ratios,
\begin{equation}
\RKp = \frac{\mathrm{BR}(B\to K^{(*)} \mu^+ \mu^- )}{\mathrm{BR}(B\to K^{(*)} e^+ e^- )}\ .
\end{equation}
As a consequence of Lepton Flavour Universality (LFU),
$R_K = R_{K^*} = 1$ in the SM. However, the latest experimental results from
LHCb, in the specified regions of $q^2$ di-lepton invariant mass, are~\cite{LHCb:2021trn,Aaij:2017vbb}:
$R_K^{[1.1, 6]} = 0.846^{+0.042}_{-0.039}{}^{+0.013}_{-0.012}\,$,
$R_{K^*}^{[0.045,1.1]} = 0.66^{+0.11}_{-0.07}\pm 0.03\,$,
$R_{K^*}^{[1.1,6]} = 0.69^{+0.11}_{-0.07}\pm 0.05$.
Therefore, sizeable violations of LFU at the
$3.1\, \sigma$ level for the $R_K$ ratio and at the $2.3\, \sigma$ level for the
$R_{K^*}$ ratio in the low-$q^2$ region and $2.4\, \sigma$ in the
central-$q^2$ region have been found.

Effective Field Theories (EFT) offer a model-independent analysis of
New Physics (NP) effects. The NP contributions at an energy scale
$\Lambda \sim 
\mathcal{O}(\mathrm{TeV})$
is described by the SMEFT Lagrangian: 
\begin{equation}
  \mathcal{L}_\mathrm{SMEFT} =
  \frac{1}{\Lambda^2}\left(\Clqo^{ijkl}\, O_{\ell q(1)}^{ijkl} +
    \Clqt^{ijkl}\,  O^{ijkl}_{\ell q(3)}   \right) \ ,
\label{eq:Lagr_SMEFT}
\end{equation}
where the dimension six operators are defined as 
$O_{\ell q(1)}^{ijkl} = (\bar{\ell}_i \gamma_\mu P_L \ell_j)(\bar{q}_k
\gamma^\mu  P_L q_l)$ and
$O_{\ell q(3)}^{ijkl}= (\bar{\ell}_i \gamma_\mu \tau^I P_L
\ell_j)(\bar{q}_k \gamma^\mu \tau^I P_L q_l)$, 
 $\ell$ and $q$ are the lepton and quark $SU(2)_L$ doublets, $\tau^I$ the Pauli matrices, and ${i,j,k,l}$ denote generation indices.  
We will restrict our analysis to operators including only third generation quarks and 
same-generation leptons: 
\begin{equation}
  \Clq^e \equiv \Clq^{1133}\ , \qquad\qquad \Clq^\mu \equiv \Clq^{2233}\ ,
  \qquad\qquad \Clq^\tau \equiv \Clq^{3333}\ .
\label{eq:wcs}
\end{equation}
Since the $\mathcal{O}_{\ell q}$ operators also produce unwanted contributions
to the $B \to K^{(*)} \nu \bar{\nu}$ decays, we will fix the relation
$\Clqo^i = \Clqt^i \equiv \Clq^i$ in order to obey these constraints.

\begin{figure}
  \centering
  \begin{tabular}{cc}
\includegraphics[width=0.5\linewidth]{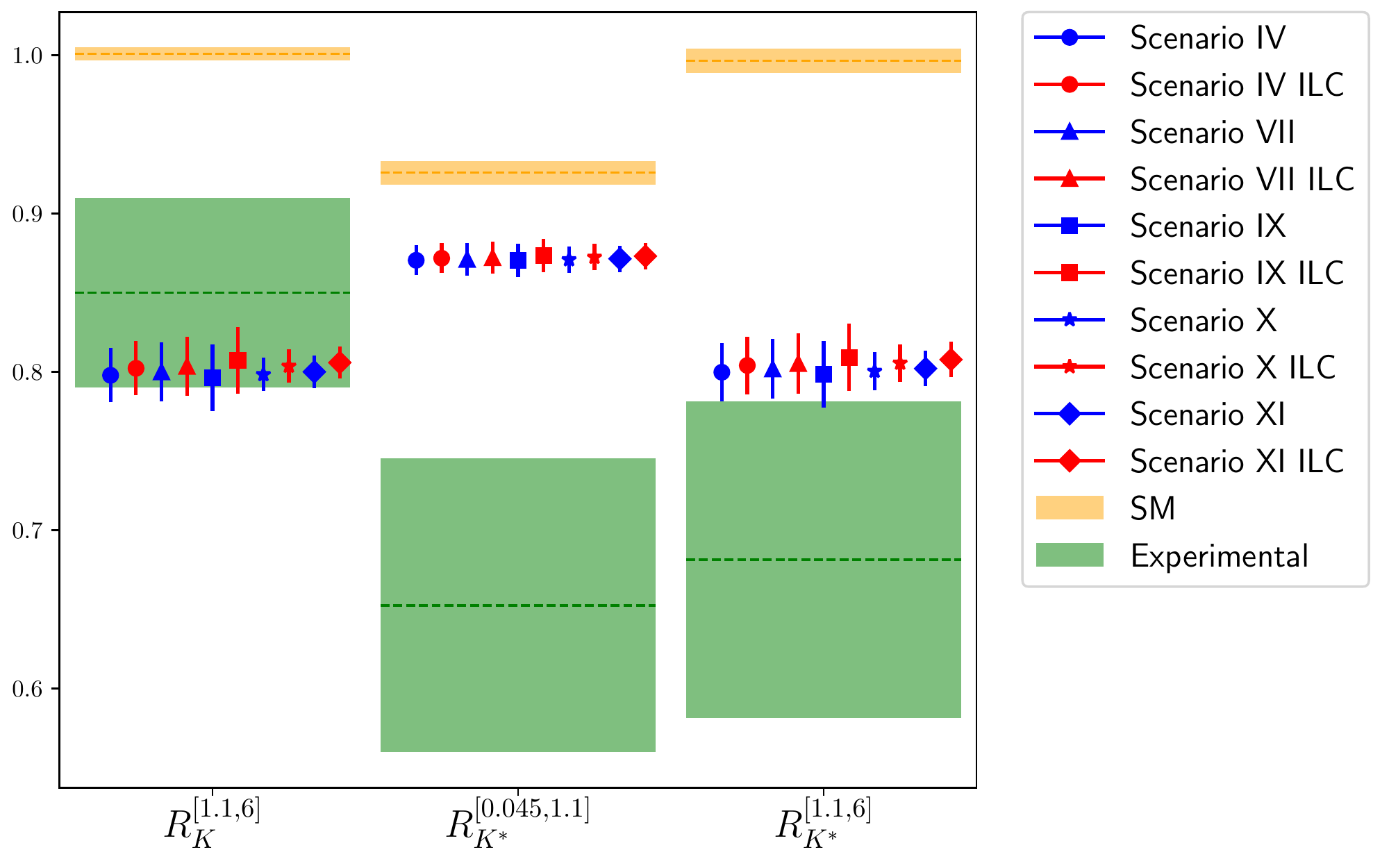}&
\includegraphics[width=0.35\linewidth]{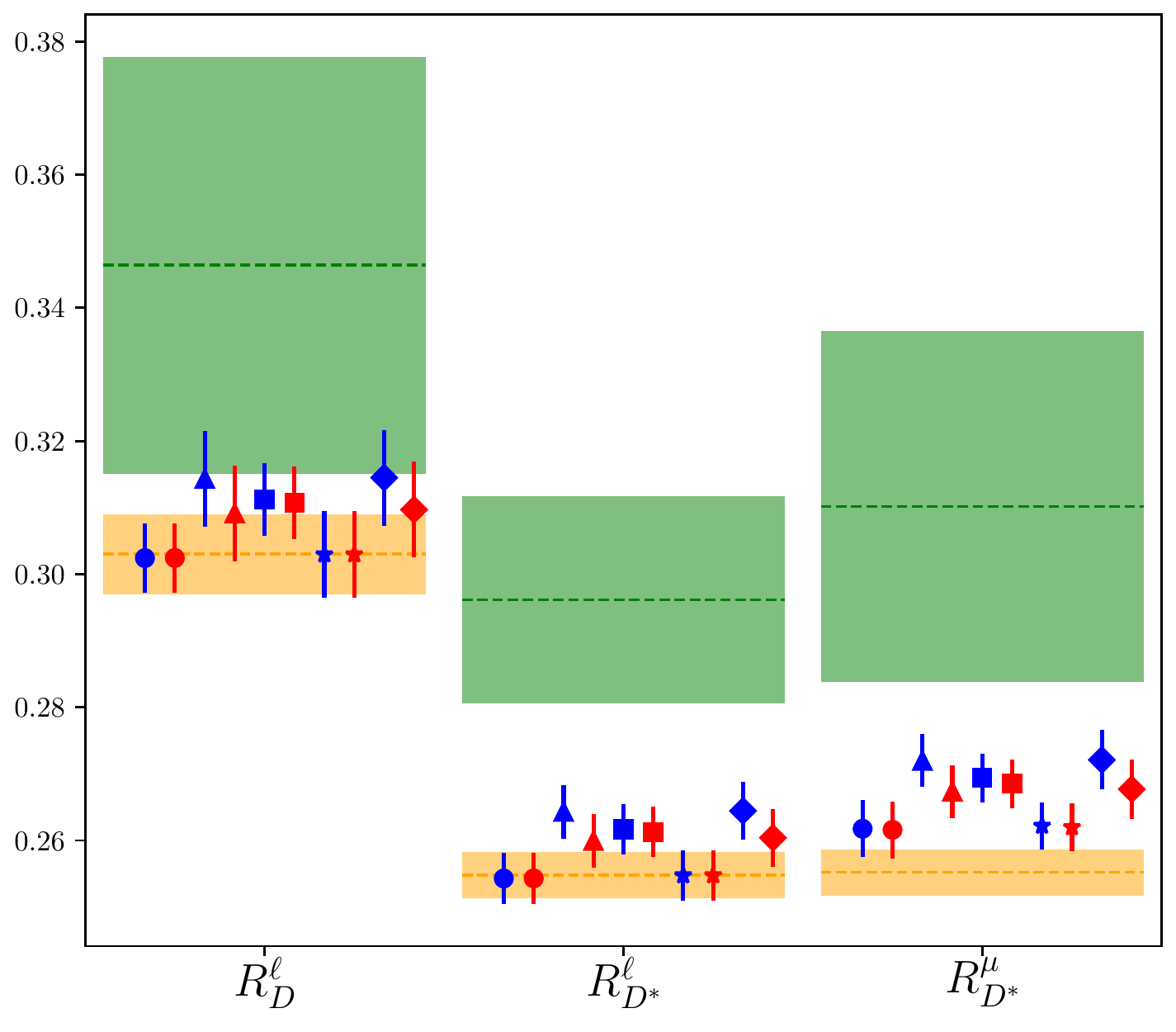}
  \end{tabular}
  \caption{Central value and $1 \sigma$ uncertainty of the $\RKp$
  observables (left), and $\RDp$ observables (right) in scenarios 
  IV, VII, IX, X and XI (blue lines for current predictions, red lines
  for ILC-based predictions), compared to the SM prediction (yellow) and
  experimental measurements (green).}
\label{fig:RkRdresults}
\end{figure}

The above effective operators affect a large number of
observables, connected between them via the Wilson coefficients.
In order to have a complete analysis of the implications of the
experimental measurements in flavour physics observables, a global fit
to the available experimental data
is required. We performed a global fit in~\cite{Alda:2018mfy,Alda:2020okk},
where an extensive list of references to previous analyses is included.
\begin{table}
    \small
  \centering
  \begin{tabular}{|c|c|c|c|c|c|c|c|}\hline
\multicolumn{2}{|c|}{\multirow{2}*{Scenario}}&\multirow{2}*{$C_{\ell
    q}^e$} &\multirow{2}*{$C_{\ell q}^\mu$}&\multirow{2}*{$C_{\ell
    q}^\tau$} &\multirow{2}*{$\Delta\chi^2_\mathrm{SM}$} & Pull & Pull
\\ \multicolumn{2}{|c|}{} & & & & & from SM & to VII\\\hline
I& $e$	& $-0.14 \pm 0.04$ & & & 8.84 &	2.97 $\sigma$ & 4.37 $\sigma$\\\hline
II& $\mu$ & & $0.10 \pm 0.04$ & & 5.47 & 2.34 $\sigma$ & 4.73 $\sigma$\\\hline
III& $\tau$ & & & $-0.38\pm0.19$ & 3.85 & 1.96	$\sigma$ & 4.89 $\sigma$\\\hline
IV& $e$ and $\mu$ & $-0.25 \pm 0.07$ &	$0.24 \pm 0.06$	& & 28.42 & 4.97
$\sigma$ & 1.75 $\sigma$\\\hline
V& $e$ and $\tau$ & $-0.14 \pm 0.06$ & & $-0.4 \pm 0.3$ & 12.98 & 3.17 $\sigma$ & 4.30 $\sigma$\\\hline
VI& $\mu$ and $\tau$ &	& $0.10 \pm 0.06$ & $-0.3 \pm 0.3$ & 8.73 & 2.49
$\sigma$ & 4.77 $\sigma$\\\hline
VII& $e$, $\mu$ and $\tau$ & $-0.25 \pm 0.02$ & $0.211 \pm 0.016$ &
$-0.3 \pm 0.4$ & 31.50 & 4.97 $\sigma$ &\\\hline
VIII& $e = \mu = \tau$	& $-0.0139 \pm 0.0003$ & $-0.0139 \pm 0.0003$ &
$-0.0139 \pm 0.0003$ & 0.30 & 0.55 $\sigma$ & 5.23 $\sigma$\\\hline
IX& $e = -\mu = \tau$ & $-0.232\pm0.001$ & $0.232\pm0.001$ &
$-0.232\pm0.001$ & 30.74 & 5.54 $\sigma$ & 0.41 $\sigma$\\\hline
X & $e = -\mu$ & $-0.23\pm 0.04$ & $0.23\pm 0.04$ & & 28.13 & 5.30 $\sigma$ & 1.32 $\sigma$\\\hline
XI & $e = -\mu$, $\tau$ & $-0.23 \pm 0.02$ & $0.23\pm0.02$ & $-0.3 \pm 0.3 $ & 30.51 & 5.17 $\sigma$ & 1.00 $\sigma$ \\\hline
  \end{tabular}
  \caption{Best fit values and pulls from the SM and of scenario VII for several combinations
  of the $\Clq^i$ operators; with one, two and three of the $C_{\ell q}$
  operators receiving NP contributions.}  
\label{tab:results}
\end{table}
In~\cite{Alda:2020okk} the global fits to the $\Clq$ coefficients
have been performed by using the package \texttt{smelli} v1.3.
This fit includes $b\to s \mu\mu$ observables; 
the branching ratio of $B_s \to \mu\mu$, the angular
observables $P_5'$ and $\RKp$, as well as $\RDp$, $b \to s \nu \bar{\nu}$ and the
electroweak (EW) precision observables ($W$ and $Z$ decay widths and branching ratios
to leptons). The goodness of each fit is evaluated with its difference of $\chi^2$
with respect to the SM value. We have defined some specific phenomenological
scenarios by choosing several combinations of the $\Clq^i$ operators~\cite{Alda:2020okk},
as shown in the first two columns of Table~\ref{tab:results}. The best fit values for the
$\Clq^i$ operators and the pulls from the SM and of scenario VII are included in this table.
Scenario VII corresponds with the more general one in which the three $\Clq$ operators
  - $\Clq^e , \Clq^\mu$ and $\Clq^\tau$ - receive independent NP
  contributions. We found that the prediction of the $\RDp$ and $\RKp$ observables in this
  scenario is improved. We also found that
an scenario within the condition $\Clq^e = -\Clq^\mu = \Clq^\tau$ imposed,
provides a similar fit goodness with a smaller set of free parameters.
In general, the best results are obtained when $\Clq^e \approx - \Clq^\mu$
(scenarios IV, VII, IX, X and XI). This indicates a maximal violation of LFU,
while lepton-universal NP is heavily constrained by the EW observables.
The results for the \RKp and \RDp observables in the best fit points
for the above scenarios are given in Figure~\ref{fig:RkRdresults}.
We remark that for the \RDp ratios, the $\Clq^\tau$ contributions
are needed, making scenarios VII, IX and XI preferred over IV and X.
Clearly, for a simultaneous analysis of
predictions for the \RDp and \RKp observables, the scenarios with
three non-universal Wilson coefficients are favoured.

\begin{figure}
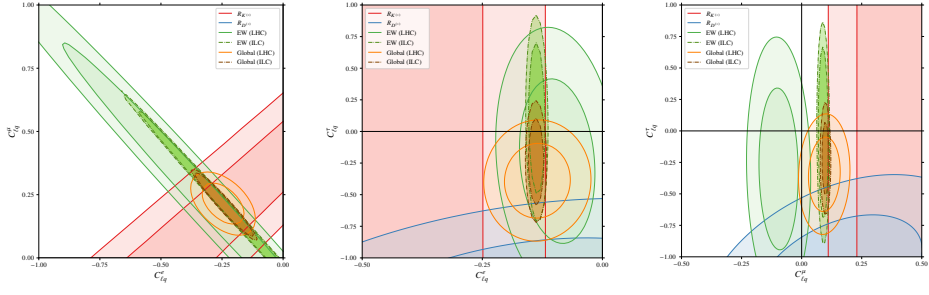

  \centering
  \begin{tabular}{ccc}
\resizebox{0.25\linewidth}{!}{\input{figures/scIV_EW_zoom.pgf}}&
\resizebox{0.25\linewidth}{!}{\input{figures/scV_EW_zoom.pgf}}&
\resizebox{0.25\linewidth}{!}{\input{figures/scVI_EW_zoom.pgf}}
\end{tabular}
\caption{$1\sigma$ and $2\sigma$ contours for several scenarios:
  Scenario IV (left),
  Scenario V (center), and Scenario VI (right). Solid
  lines correspond to the current fits, dash-dotted lines to the fits including the ILC
  projections.}
\label{fig:fitresults}
\end{figure}

Let us stress that new experimental inputs should provide valuable
new information to cast light on B anomalies. In this concern,
the future $e^+e^-$ linear collider ILC will offer the opportunity to
use the increased
precision of the EW observables, thanks to the huge production of
$Z$ bosons, to further constrain the global fits.
In order to asses the impact of the improved precision
on our analysis, we have performed a new global
fit in~\cite{Alda:2021ruz}, by using for the central
values of the EW observables their predictions as in our
previous work~\cite{Alda:2020okk} and taking the uncertainty from the ILC at
$\sqrt{s} =$ 250 GeV projections from~\cite{Strategy:2019vxc}.
For comparison, the results of the fits to scenarios IV, V and VI; in which two 
of the Wilson coefficients receive NP contributions simultaneously, both for the
current fits at LHC and for the ILC projections, are presented in Figure~\ref{fig:fitresults}.
One can conclude that the LFU-conserving direction of the fit,
corresponding to the linear combination $\Clq^e + \Clq^\mu$, is even more tightly
constrained due to the better precision of the EW observables. The LFUV direction of the fit
remains unchanged, since the EW observables are not sensitive to these deviations. 

The results for the central value and $1 \sigma$ uncertainty
for the \RKp and \RDp observables in the best fit points
for Scenarios IV, VII, IX, X and XI at ILC are also included in Figure~\ref{fig:RkRdresults}.
The error of the $\RKp$ observables is improved
up to factor of 3. When including the ILC projections, the error in all
those observables is dominated by the theoretical
uncertainty, due to that the allowed region for the Wilson coefficients in the fits is reduced.

Summarising, it is difficult to find an easy common explanation for all
flavour anomalies, but the analysis of the effects of the global fit to the Wilson
coefficients is mandatory. The anomalies can be described
by NP displaying a maximal violation of universality between electrons
and muons, but some NP in the tau sector is also needed. It is clear
that new experimental inputs and updated global fits to date are
  needed to clarify the present situation.


\begin{thebibliography}{99}
\bibitem{Amhis:2019ckw}
  Y.~S.~Amhis {\it et al.} [HFLAV], 
  Eur.\ Phys.\ J.\ C {\bf 81} (2021) 
  226
  [arXiv:1909.12524 [hep-ex]].
  
\bibitem{LHCb:2021trn}
R.~Aaij \textit{et al.} [LHCb],
[arXiv:2103.11769 [hep-ex]].

\bibitem{Aaij:2017vbb}
  R.~Aaij {\it et al.} [LHCb], 
  JHEP {\bf 1708} (2017) 055
  [arXiv:1705.05802 [hep-ex]].

  \bibitem{Alda:2018mfy}
    J.~Alda, J.~Guasch, S.~Pe\~naranda,
    {Eur. Phys. J. C \textbf{79} (2019) 
      588}
    [arXiv:1805.03636 [hep-ph]]

\bibitem{Alda:2020okk}
  J.~Alda, J.~Guasch, S.~Pe\~naranda,
{[arXiv:2012.14799 [hep-ph]]}.
  
\bibitem{Alda:2021ruz}
J.~Alda, J.~Guasch, S.~Pe\~naranda,
[arXiv:2105.05095 [hep-ph]].

\bibitem{Strategy:2019vxc}
R.~K.~Ellis \textit{et al.},
[arXiv:1910.11775 [hep-ex]].

\end{thebibliography}
\end{document}